\documentclass{aa}
\usepackage{graphicx,epsf,epsfig}

\usepackage{txfonts}
\begin{document}
   \title{The long period Intermediate Polar 1RXS\,J154814.5-452845\thanks{Based on observations collected at the European Southern 
Observatory, La Silla, Chile. Program: 71.D-0195}
}

   \author{D. de Martino
          \inst{1}
          \and
          J.-M. Bonnet-Bidaud\inst{2}\and M. Mouchet\inst{3}\and
B.T. G\"ansicke\inst{4}
\and F. Haberl\inst{5} \and C. Motch\inst{6}
}

\offprints{D. de Martino}

   \institute{INAF--Osservatorio Astronomico di Capodimonte, Via
Moiariello 16, I-80131 Napoli, Italy\\
              \email{demartino@na.astro.it}
         \and
Service d'Astrophysique, DSM/DAPNIA/SAp, CE Saclay, F-91191 Gif sur Yvette
Cedex, France\\
\email{bonnetbidaud@cea.fr}
\and
APC, UMR 7164, University Denis Diderot, 2 place Jussieu, F-75005 and 
LUTH, Observatoire de Paris, Section  de Meudon, F-92190 Meudon, 
France\\
\email{martine.mouchet@obspm.fr} 
\and
Department of Physics, University of Warwick, Coventry CV4 7AL,
UK\\ 
\email{ boris.gaensicke@warwick.ac.uk}
\and
Max-Planck-Instit\"ut f\"ur Extraterrestrische Physik,
Giessenbachstra{\ss}e, Postfach 1312, 85741 Garching, Germany \\
\email{fwh@mpe.mpg.de}
\and
Observatoire Astronomique, UA 1280 CNRS, 11 rue de l'Universit\'e, 67000 
Strasbourg, France\\
\email{motch@astro.u-strasbg.fr}
             }

   \date{Received 2005 07, 21; accepted 2005 12, 18}

   \abstract{We present the first time resolved medium resolution optical 
spectroscopy of the 
recently identified peculiar Intermediate Polar (IP) \object
1RXS\,J154814.5-452845, which 
allows  us to precisely determine the binary orbital period ($\rm 
P_{\Omega}=9.87\pm$0.03\,hr) and the white dwarf spin period 
($\rm 
P_{\omega}=693.01\pm$0.06\,s). This system is then the third  just 
outside the  purported $\sim$6-10\,hr IP orbital period gap and the 
fifth of the small 
group of long period
IPs, which has a relatively high degree of asynchronism. 
From the presence of weak red absorption features, we 
identify the secondary star with a spectral type  K2$\pm$2\,V, which 
 appears to have evolved on the nuclear timescale.
From the  orbital radial 
velocities of emission and the red absorption lines a mass ratio 
$q=0.65\pm0.12$ is found. The  
masses of the components are estimated to be $\rm 
M_{WD} \geq 0.5 \,M_{\odot}$ and $\rm 
M_{sec}=0.4 -0.79\,\,M_{\odot}$ and 
the binary inclination  $25^o < i \leq 58^o$. A distance between 
540-840\,pc 
is estimated. At this distance, the  presence of peculiar absorption 
features surrounding Balmer emissions cannot be due to the contribution of 
the white dwarf photosphere and their spin modulation suggests an
origin in the magnetically confined
accretion flow. The white dwarf  is also not accreting at a 
particularly high rate ($\rm \dot M < 5\times 10^{16}\,g\,s^{-1}$), for its 
orbital period.
The  spin-to-orbit period ratio $\rm P_{\omega}/P_{\Omega}$=0.02 and
the low mass accretion rate suggest that  this system is 
 far from spin equilibrium. The magnetic moment of the accreting white 
dwarf is found  to be  $\mu < 4.1\times10^{32}$\,G\,cm$^3$,
indicating a low magnetic field system.

   }

   \maketitle
%

\section{Introduction}

Intermediate Polars (IP) are magnetic  
Cataclysmic Variables (mCVs) characterized by strong and hard 
($\rm kT \sim$ 15\,keV)   X-ray pulses usually  
at the rotational period of the accreting white dwarf (WD) 
 ($\rm P_{\omega} << P_{\Omega}$). The X-ray 
pulsations indicate
that the accretion flow is magnetically channelled towards the WD polar 
regions. However, the lack of detectable optical-IR circular polarization 
in  most of these systems prevents the measure of the WD magnetic 
field.   It 
indicates  that the magnetic field of the accreting WD is lower 
(B$<$5-10\,MG) than  that detected in the strongly magnetized 
(B$\sim$10-230\,MG) synchronous ($\rm P_{\omega} = P_{\Omega}$) polar 
systems 
(see Warner \cite{Warner95} for a comprehensive review of mCVs).   
The orbital periods of IPs are generally long ($\rm P_{orb}>$4\,hr), with 
only three systems confirmed to be IPs by X-rays 
below the  so-called 2-3\,hr orbital period gap,  
whilst the polars are typically found at shorter ($<$4\,hr) orbital 
periods,
with most systems below the period gap.  
The different location of the two subclasses in the orbital 
period distribution of mCVs suggests that IPs 
will synchronize in their evolution path towards  short orbital 
periods and hence they might be progenitors of  polar systems (Norton et 
al. \cite{norton04}). However, the difference in the 
magnetic field strengths was the prime reason 
for rejecting the hypothesis of IPs evolving into polars, although  recent 
works on magnetic field evolution in accreting WD might explain the 
discrepancy  (Cumming \cite{Cumming02}). 
The  large variety of  observational properties found in the  
40 or so systems known to date, still  need to 
be understood in terms of accretion and evolutionary state. In particular 
a wide  range of asynchronism seems to characterize this class
(Woudt \& Warner \cite{woudtwarner}). Such a wide range was recently 
discussed by Norton et al. 
(\cite{norton04}) in terms of spin equilibrium for magnetic accretion, 
which can take place on a wide variety of ways ranging from 
magnetized  accretion 
streams to extended accretion discs. Also, in a few systems the presence of a 
"soft X-ray" emission component, similar  to that observed in the polars 
(Buckley \cite{Buckley}) raised  the evolutionary question of 
whether these soft X-ray IPs are the  true progenitors of the 
polars. Furthermore the apparent lack of IPs in the range of orbital periods 
$\sim$6-10\,hr (the so-called "IP period gap") was only recently noticed 
(Schenker et al.  \cite{schenker}), with only one recently 
discovered system at 7.2\,hr 
(Bonnet-Bidaud et al. \cite{bonnet-bidaud05}) and 
with a handful of long period IPs (Buckley \cite{Buckley}; 
G\"ansicke et al. \cite{gaensicke05}), being the most peculiar ever known.
Hence, our understanding of 
the evolutionary  relationships among mCVs and in particular of IPs 
is still very 
poor and the addition of new systems and the study of their
properties has a great  potential to alleviate this problem.\\

The X-ray source \object 1RXS\,J154814.5-452845 (henceforth RX\,J1548) was 
identified as an  Intermediate Polar (IP) by Haberl et al. 
(\cite{Haberl02}) using optical and X-ray observations. The
693\,s  WD rotational period  was identified in both optical 
and X-ray light curves, whilst sparse optical spectroscopy and 
photometric data did not allow an unambiguous identification of the 
orbital period. The two possible 
values of 9.37$\pm$0.69\,hr or 6.72$\pm$0.32\,hr, would add 
RX\,J1548 to the 
small group of
IPs with long orbital periods, both periods locating it in the IP period 
gap. 
Furthermore, RX\,J1548 possesses a highly absorbed 
 hot black-body soft X-ray component similar to that observed in the 
IP V2400\,Oph, but different from that found in the soft IPs 
(PQ\,Gem, V405\,Aur) (de Martino et al. \cite{demartino04}) 
which instead do not suffer from strong absorption.
RXJ\,1548 is also peculiar in  that its optical spectrum, 
showing broad  absorption features underneath Balmer emissions, is 
similar to those observed  in IP V\,709 Cas 
(Bonnet-Bidaud et al 
\cite{Bonnet-Bidaud01}), which might suggest the 
contribution of the 
underlying accreting WD atmosphere (Haberl et al. \cite{Haberl02}).

\noindent In this work we present extensive high temporal resolution 
optical spectroscopy  aiming to determine the true orbital period, to 
elucidate the nature of the peculiar
absorption features and that of the stellar components of this binary.

   \begin{table}[t!]
      \caption{Summary of the observations of RX\,J1548.}
         \label{obslog}
     \centering
\begin{tabular}{c c c c c }
            \hline \hline
            \noalign{\smallskip}
Date &  Start &  Spectral  Range     & $\rm T_{expo}$ & N. Spectra\\
 UT    &  UT   &  $\AA$         & sec & \\
 \noalign{\smallskip}
            \hline
            \noalign{\smallskip}
May 27, 2003 & 01:38:32 & 4038 - 7020 & 90 & 92  \\
May 27, 2003 & 02:38:26 & 5900 - 8600 & 200 & 4  \\
May 28, 2003 & 00:10:56 & 4038 - 7020 & 90 & 160 \\
May 28, 2003 & 01:26:21 & 5900 - 8600 & 200 & 7 \\
                 \noalign{\smallskip}
            \hline
\end{tabular}
\end{table}


\section{The observations}

RX\,J1548 was observed at the NTT (ESO) telescope on May 27-28, 2003
equipped with the ESO Multi-Mode Instrument (EMMI) in low dispersion mode 
(RILD). Long slit spectra were acquired 
in sequences of typically 20  exposures each of  90\,s  with 
Grism 5 
(4038-7020 $\AA$;  FWHM resolution: 5.8$\AA$), alternating red spectra 
acquired with  Grism 6  (5900 - 8600$\AA$; FWHM resolution: 6.0$\AA$) 
with exposure time of 200\,s  after each sequence. A slit width of 
1.5'' was used for both grisms.  
A total of 11 red and 252 blue spectra were obtained. He-Ar lamps were 
acquired several times before and after each sequence to allow a check 
of wavelength calibration. The seeing during both nights was rather variable
between 0.8'' and 1.5'' and cirrus was present during the first half 
of the  first  night.  A few late type template stars in the range between 
K2V to  M4V were also  observed with  Grism 6.

Standard reduction was performed using the ESO-MIDAS package, including 
cosmic ray removal, bias subtraction, flat-field correction and 
wavelength 
calibration. The wavelength calibration was found to be accurate to 
0.2$\AA$ and 0.1$\AA$ in the blue and red ranges respectively. 
Heliocentric  corrections were applied to all radial velocities and times 
of mid-exposure.
Furthermore flux calibration was performed using the spectrophotometric 
standard EG\,274. 
The log of the observations is reported in 
Table~\ref{obslog}.

   \begin{figure}
   \centering
\includegraphics[height=8.cm,width=8.cm]{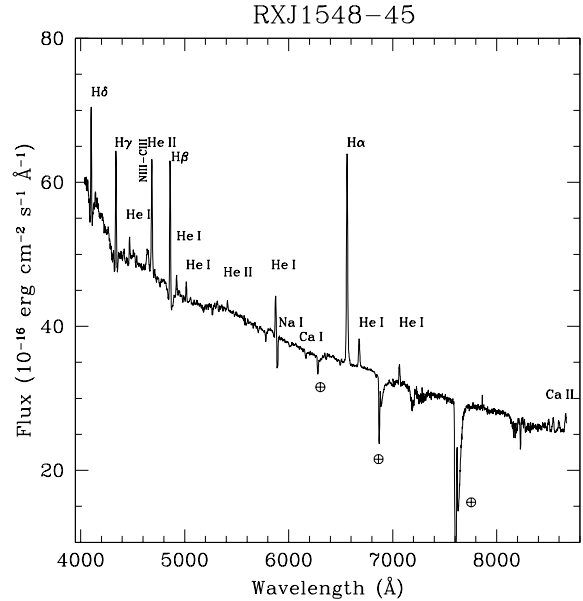}
\caption{The grand average optical spectrum of RX\,J1548.}\label{fig1}
\end{figure}

   \begin{figure*}[t!]
   \centering
\mbox{\epsfxsize=8cm\epsfysize=9cm\epsfbox{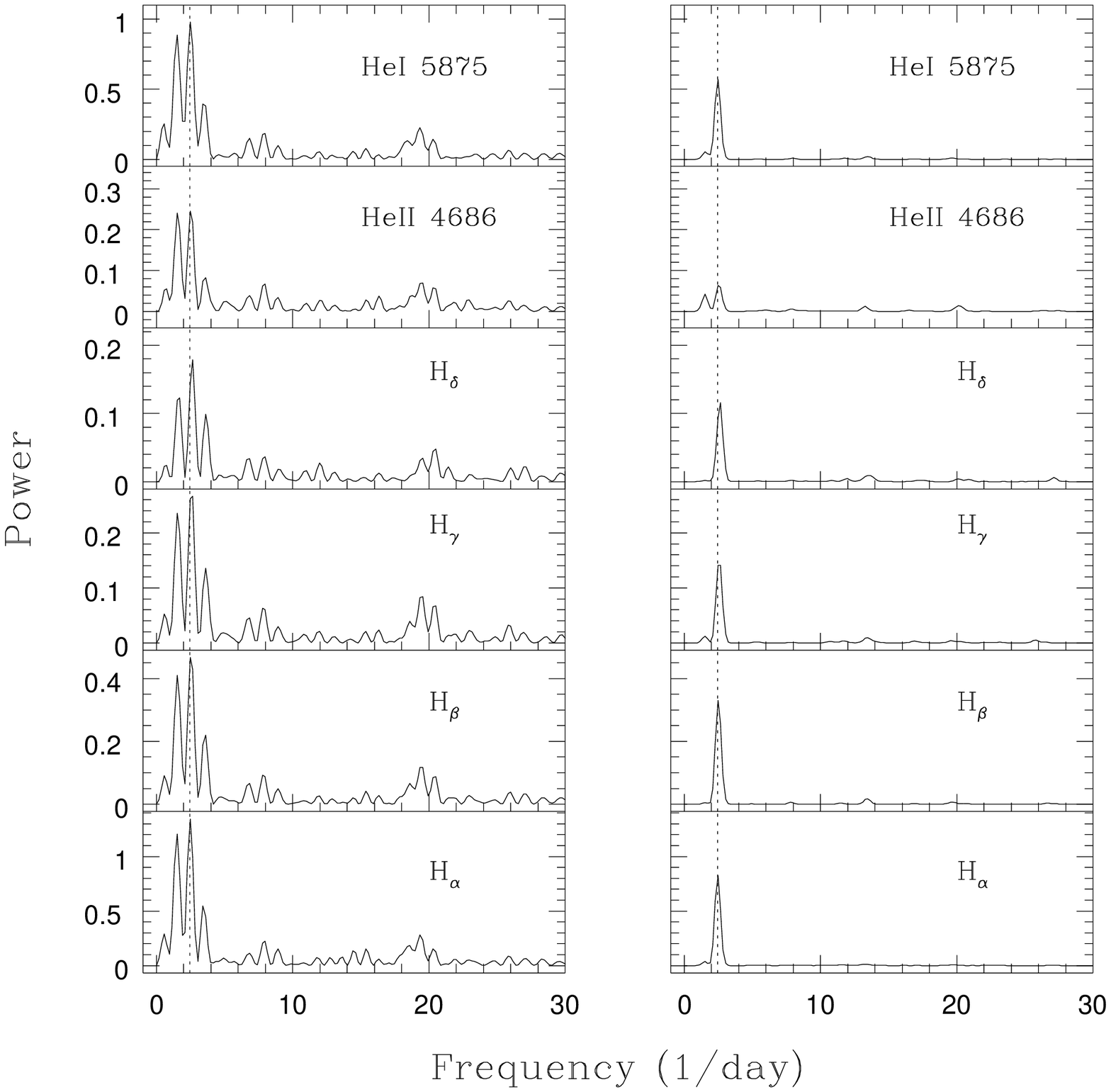}
\epsfxsize=8.cm\epsfysize=9cm\epsfbox{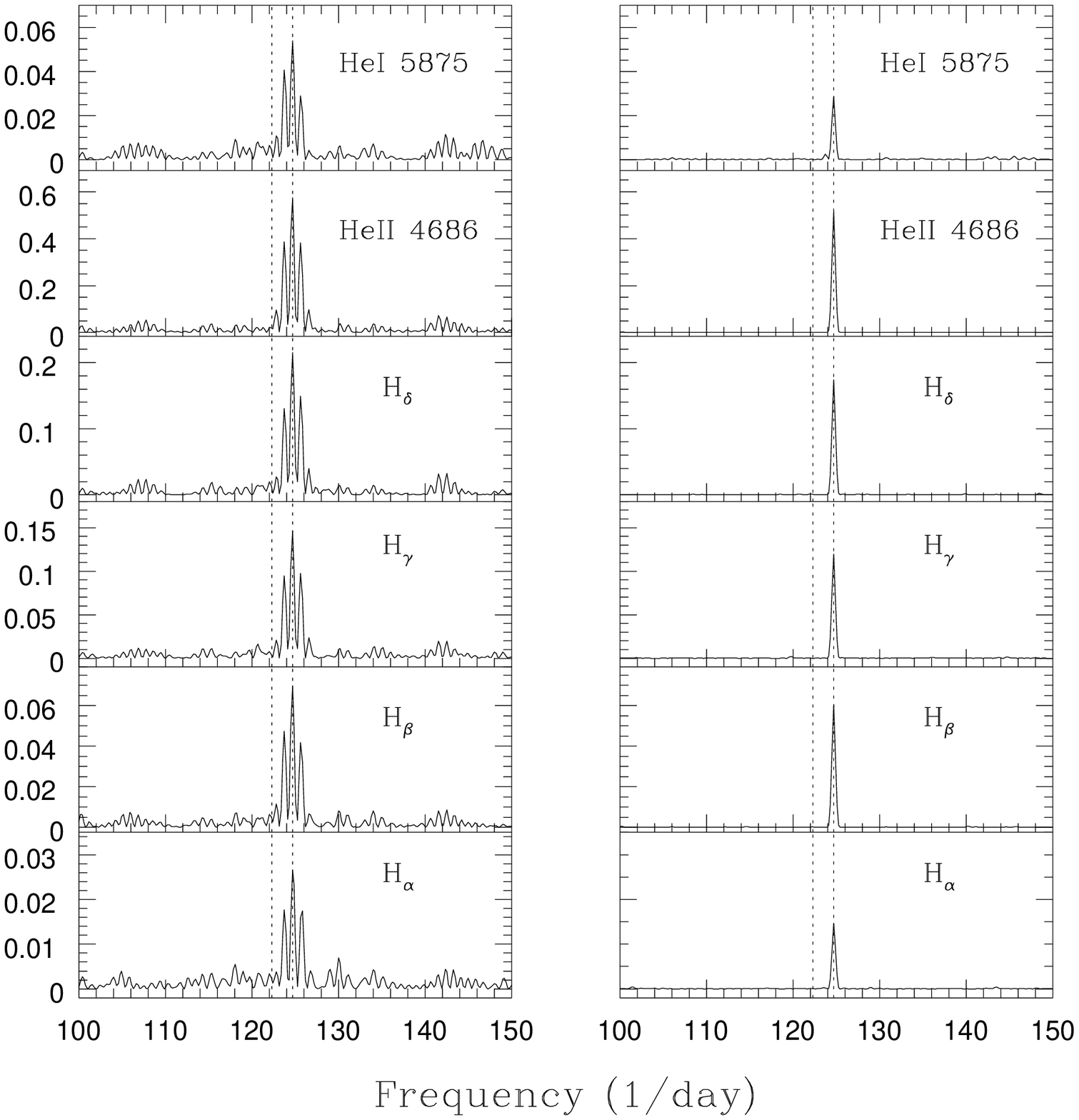}}
\caption{  Enlargements around the orbital ({\em left}) and the spin 
({\em right}) frequencies of the DFTs of radial velocities of 
emission lines (Balmer, He\,II and He\,I) and their 
CLEANED power spectra (at their
right side). Vertical dotted lines from left to right mark the positions of 
the orbital, beat and spin frequencies. }\label{fig2}
    \end{figure*}

   \begin{table*}
      \caption{Line parameters in RX\,J1548. 
}
         \label{parlines}
     \centering
\begin{tabular}{l c c c c c c }
            \hline \hline
            \noalign{\smallskip}
Line &  EW$_{\rm abs}$ &  FWHM$_{\rm abs}$ & FWZI$_{\rm abs}$  &  
EW$_{\rm em}$ &  FWHM$_{\rm em}$ & FWZI$_{\rm em}$ \\
     &  $\AA$ & $\AA$ &  km\,s$^{-1}$ &  $\AA$ & $\AA$ &  km\,s$^{-1}$ \\
 \noalign{\smallskip}
            \hline
            \noalign{\smallskip}
H$_{\delta}$ & 5.7(3) & 36(3) & 5200 & 5.0(2) &  11(1) & 1540  \\
H$_{\gamma}$ & 5.9(4) & 67(7) & 5400 & 4.5(3) & 12(1)  & 2800 \\
H$_{\beta}$  & 3.8(2) & 53(8) & 5600 & 5.8(2) & 10(1)  & 2700 \\
H$_{\alpha}$ &        &       &      & 11.4(4) & 13 (2) & 2200 \\
He\,II $\lambda$ 4686 & &     &      & 4.6(3) & 14(2)  & 2200\\
He\,I $\lambda$ 4471  & &     &      & 0.5(1) & 8(1)  & 1700\\
He\,I $\lambda$ 5875  & &     &      & 1.9(2) & 11(1) & 1500\\     
     \noalign{\smallskip}
            \hline
\end{tabular}
\end{table*}


\section{The mean optical spectrum }

The  average optical spectrum from 4038$\AA$ to 8600$\AA$, in  
Fig.\,\ref{fig1} shows  the typical emission lines of Balmer, He\,I 
($\lambda \lambda$ 7065, 6678, 5875, 5016, 4921 and 4471), He\,II 
($\lambda \lambda$ 4686 and 5411) as well as
the C\,III/N\,III blend ($\lambda \lambda$ 4640-4650) and the
Ca\,II  triplet ($\lambda \lambda$ 8498, 8542 and 8662). 
The average flux  level (V$\sim$15\,mag) was  
0.4\,mag fainter  than when observed in 1998 by Haberl et al. 
(\cite{Haberl02}). 
The previously noticed broad absorption features around 
Balmer emissions  have decreasing 
depth from H$_{\delta}$ to H$_{\beta}$ and are absent around H$_{\alpha}$ 
and Helium lines. This could suggest a  contribution from the 
underlying WD atmosphere. 
 In Table\, \ref{parlines} we report the measured 
parameters of the main lines obtained by Gaussian fits. For the 
Balmer lines we used two Gaussians to account for the absorption and 
emission components. The FWHMs of absorptions range between
35-70\,$\AA$. When compared to those measured from DA WD model atmospheres 
(Koester et al. \cite{koester}), they do not provide useful constraint on 
temperature and gravity especially those of H$_{\delta}$ and H$_{\beta}$ lines 
which have distorted profiles, possibly due to close-by weak emissions. 
Also, WD model atmosphere fits even for the more symmetric H$_{\gamma}$ 
absorption  profile (once the emission component is removed), the 
WD atmosphere models give a very
poor match  especially in the line wings, thus making the WD 
identification unsecure.
As it will be shown in sect.\,4.2 and discussed in 
sect.\,5.1, a WD origin is furthermore ruled out. Hence, an alternative
origin in the accretion flow close to the WD surface is proposed in 
sect.\,5.1.

The red portion of the spectrum presents weak absorptions identified 
as
Ca\,I ($\lambda \lambda$ 5270 and 6162), Na\,I ($\lambda$ 5889)
Mg\,b at 5170\,$\AA$ and  metallic lines at 6497\,$\AA$. TiO 
bands and CaH at  6386\,$\AA$
are absent and the Na\,I  near-infrared doublet (($\lambda 
\lambda$ 8183, 8195)  is weak and consistent with telluric
absorption as well as other strong absorptions in this part of the spectrum
(Fig.\,\ref{fig1}).  Following Torres-Dodgen \& Weaver (\cite{torresdodgen})
classification, these characteristics indicate a main sequence 
K type star but earlier than K6, and  the absence of Ca\,I ($\lambda$ 
6122) is typical of K0-K3 stars.

   \begin{figure*}[t!]
   \centering
\mbox{\epsfxsize=8cm\epsfysize=9cm\epsfbox{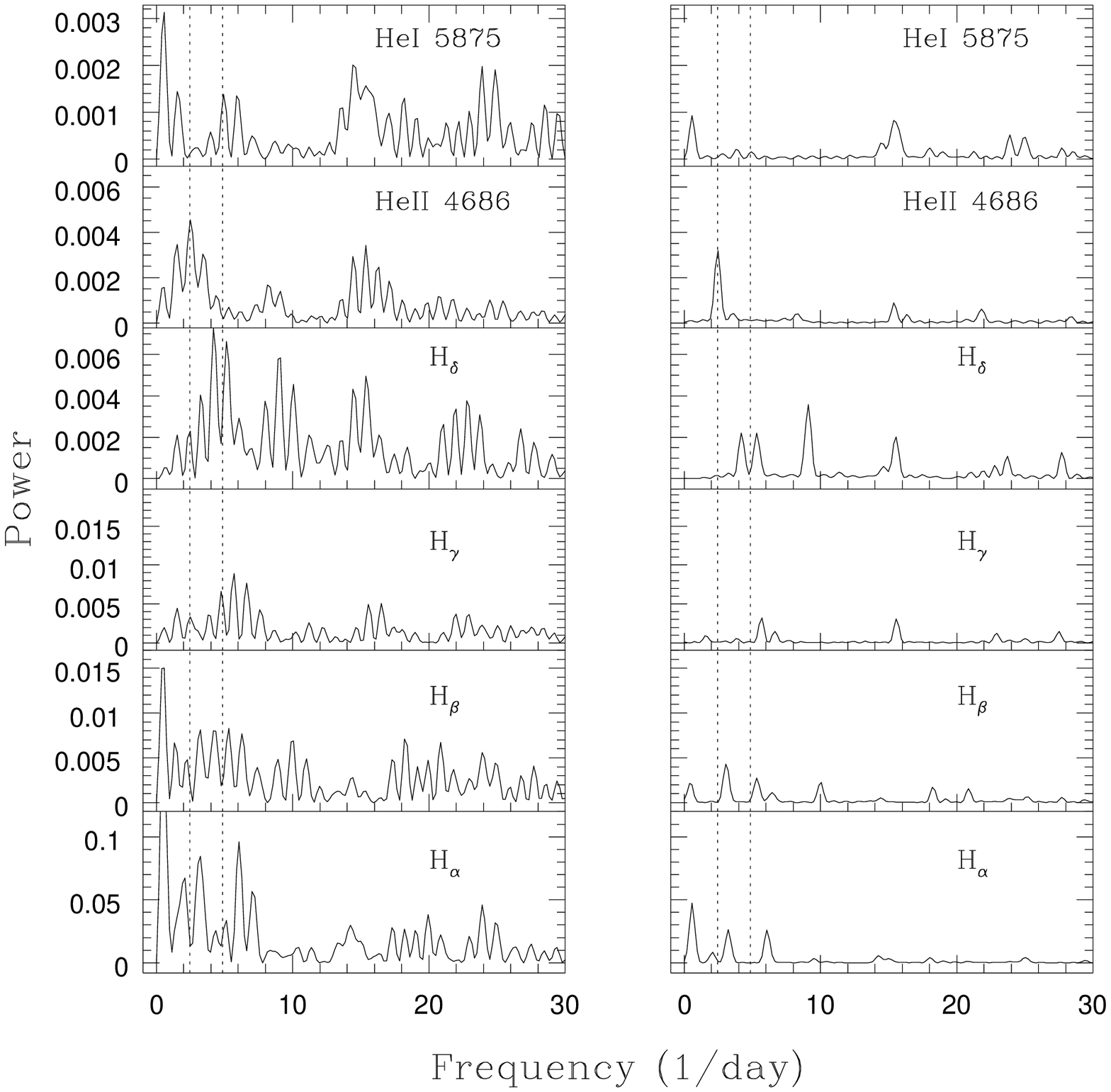}
\epsfxsize=8.cm\epsfysize=9cm\epsfbox{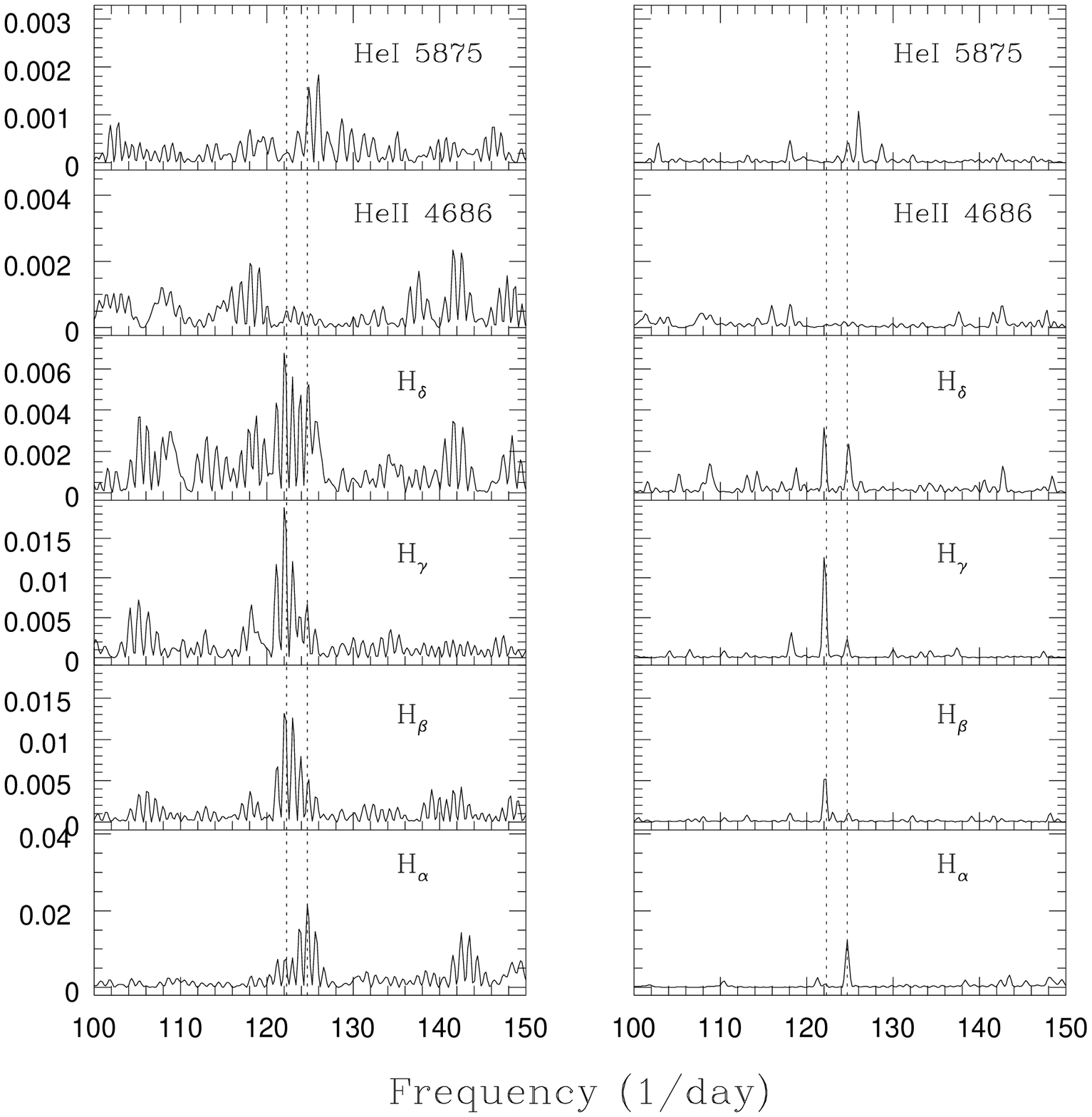}}
\caption{ Enlargements around the orbital ({\em left}) and the spin 
({\em right}) frequencies of the DFTs of EWs of emission lines 
(Balmer, He\,II and He\,I) and their CLEANED power spectra 
(at their right side). The  positions of 
the orbital, its first harmonic, the beat and spin frequencies are also 
marked with vertical 
dotted lines. }\label{fig3}
    \end{figure*}

%
\begin{figure}[h!]
\centering
\includegraphics[width=8cm]{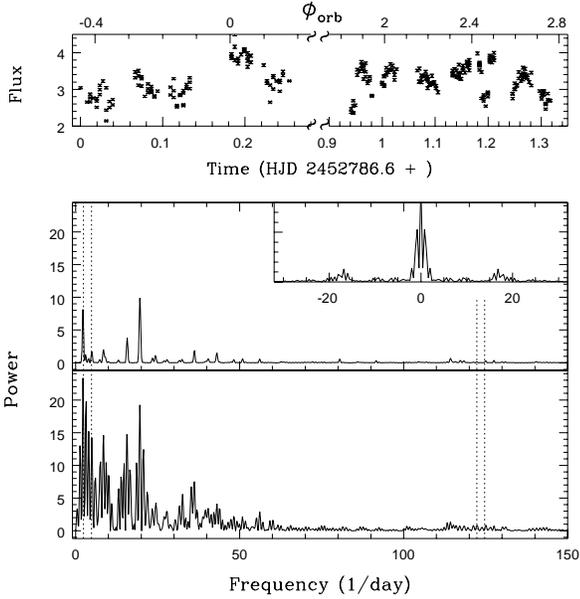}
\caption{ {\em Upper panel:}  The continuum flux light curve
measured in  line-free regions of the spectrum from 4040 to 6800\,$\AA$.  
{\em  Lower panel:} The DFT of the continuum light curve. {\em  Middle 
panel:} The CLEANED power spectrum of the same time series.  
Note the absence of signal at the spin and beat frequency and the 
presence of power at the orbital frequency. 
The spectral window of the data is shown in the inserted panel. The peaks 
at $\sim$20\,day$^{-1}$, partially due to the spectral window and 
hence partially removed by the CLEAN algorithm, are
likely due to slit losses (see text).} 
\label{fig4}
\end{figure}


\section{Timing analysis}

In order to inspect emission line variability a Gaussian
fit was performed to the Balmer and Helium emission lines of each
blue spectrum. 
The measured radial velocities and equivalent widths (EWs) show a clear 
short term and 
long term variability. These were then 
Fourier analysed (excluding 26 spectra at the beginning of the second
night as discussed in sect.\,4.1) and 
their DFTs (Discrete Fourier Transforms) are shown in Fig.\,\ref
{fig2}  and Fig.\,\ref {fig3}. We have furthermore used the 
CLEAN algorithm (Roberts et al. \cite{robertsetal}) to remove the 
windowing effects in the data which are 
also visible in the spectral 
window shown in Fig.\,\ref {fig4}.  The radial 
velocity power spectrum,
once cleaned from windowing, clearly 
reveals  two peaks at $\omega$=124.6\,d$^{-1}$ and at
$\Omega$=2.43\,d$^{-1}$ (Fig.\,\ref{fig2}). The absence of 
the beat $\omega - \Omega$= 122.17\,d$^{-1}$ signal confirms  the former 
as the rotational frequency of the white dwarf and the latter as the 
binary orbital frequency.
A trend in amplitudes is observed in the Balmer lines, 
H$_{\alpha}$ has the  strongest 
peak at the orbital frequency and the weakest peak at the spin frequency,
whilst H$_{\delta}$ has the strongest peak at the spin and the weakest  
at the orbital frequency. He\,II ($\lambda$ 4686) is also strongly 
modulated at the spin frequency while He\,I lines, at $\lambda \lambda$ 
4471 (not shown in Fig.\,\ref {fig2}) and 5875,  show the strongest 
modulation at the orbital period,  similarly to H$_{\alpha}$, 
suggesting a common line forming region.
EWs of the same lines were also Fourier analysed and are shown
in  Fig.\,\ref {fig3} together with their 
CLEANED  spectra. The power spectra are much noiser, but reveal that the 
spin 
frequency is  present in H$_{\alpha}$ and H$_{\delta}$ and weakly 
present in H$_{\beta}$ and H$_{\gamma}$, these latter instead show a much
stronger signal at the beat $\omega - \Omega$ frequency. On the other
hand the spin and beat modulations are absent in He\,II, while He\,I
behaves similarly to H$_{\alpha}$ as also seen in the radial velocities.  
The orbital modulation is clearly present in the EWs of He\,II line but
not in He\,I, whilst in Balmer lines an excess of power is present at 
low  frequencies.
 While the CLEANED spectra  might suggest a 
variability of the EWs of Balmer emission lines at the fundamental and 
first harmonic of the orbital 
frequency with a possible anti-correlated behaviour in their strengths   
moving from H$_{\alpha}$  to H$_{\delta}$, the significance of the peaks
in the low frequency region is below the 3$\sigma$.

  To study the periodic variability in the Balmer absorption 
components, we also performed a fit to the Balmer lines using 
 two Gaussians in order to reproduce  the emission and the absorption
components in each blue spectrum. Due to the 
low signal-to-noise of the individual spectra the fitting procedure failed 
to reproduce the absorption in $\sim 35\%$ of the cases, thus preventing 
us to detect a periodic signal in the power spectra of both radial 
velocities and EWs of the absorption components. However, as it will be 
shown below, the spin and orbital variability can be instead detected 
once the spectra are binned and phased at these periods.

 The continuum was also  inspected for periodic variability  by 
performing Fourier analysis on the flux light curve of 
 line-free  regions of the spectrum between 4040 and 6800\,$\AA$ 
(Fig.\,\ref{fig4}). 
In contrast to the photometric results by Haberl et al. 
(\cite{Haberl02}), neither the spin nor the beat modulations are
detected. The  orbital variability appears instead in the CLEANED 
spectrum, though other peaks in the low frequency portion are present.
Similar power spectra are obtained using the whole 4038-7020\,$\AA$ range 
without excluding emission lines.  Since only part of the observed
signal in the low frequency portion is due to the spectral window 
(shown in Fig.\,4) and
as seen in the light curve the spectrophotometry  might  be 
affected by the variable seeing  which might introduce spurious low 
frequency behaviour, we  limit ourselves to report on the absence of high 
frequency variations in the continuum with an upper limit of $\sim 1\%$ 
for the spin modulation.

   \begin{figure*}[t!]
   \centering
\mbox{\epsfxsize=8cm\epsfysize=12cm\epsfbox{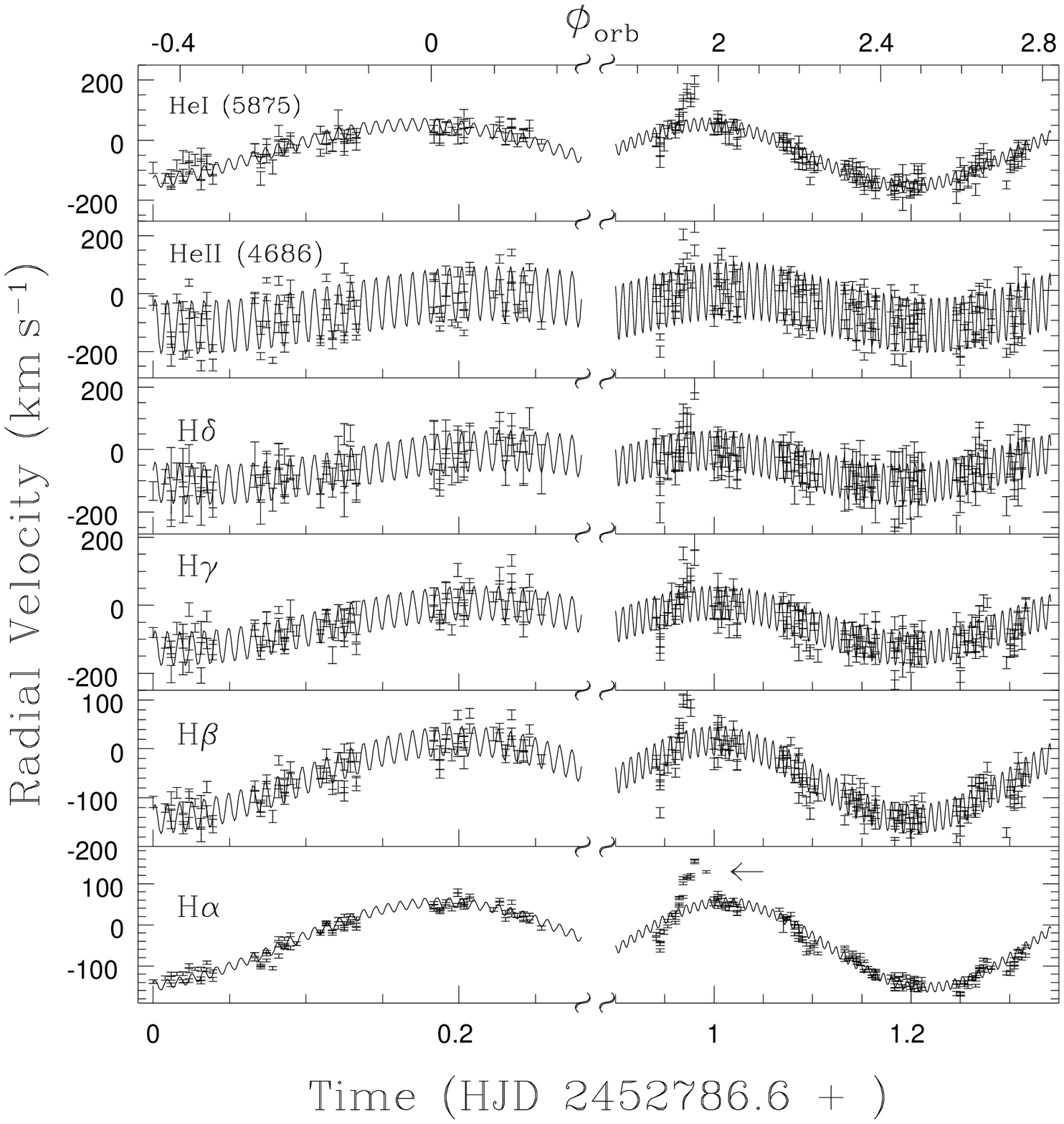}
\epsfxsize=8.cm\epsfysize=12cm\epsfbox{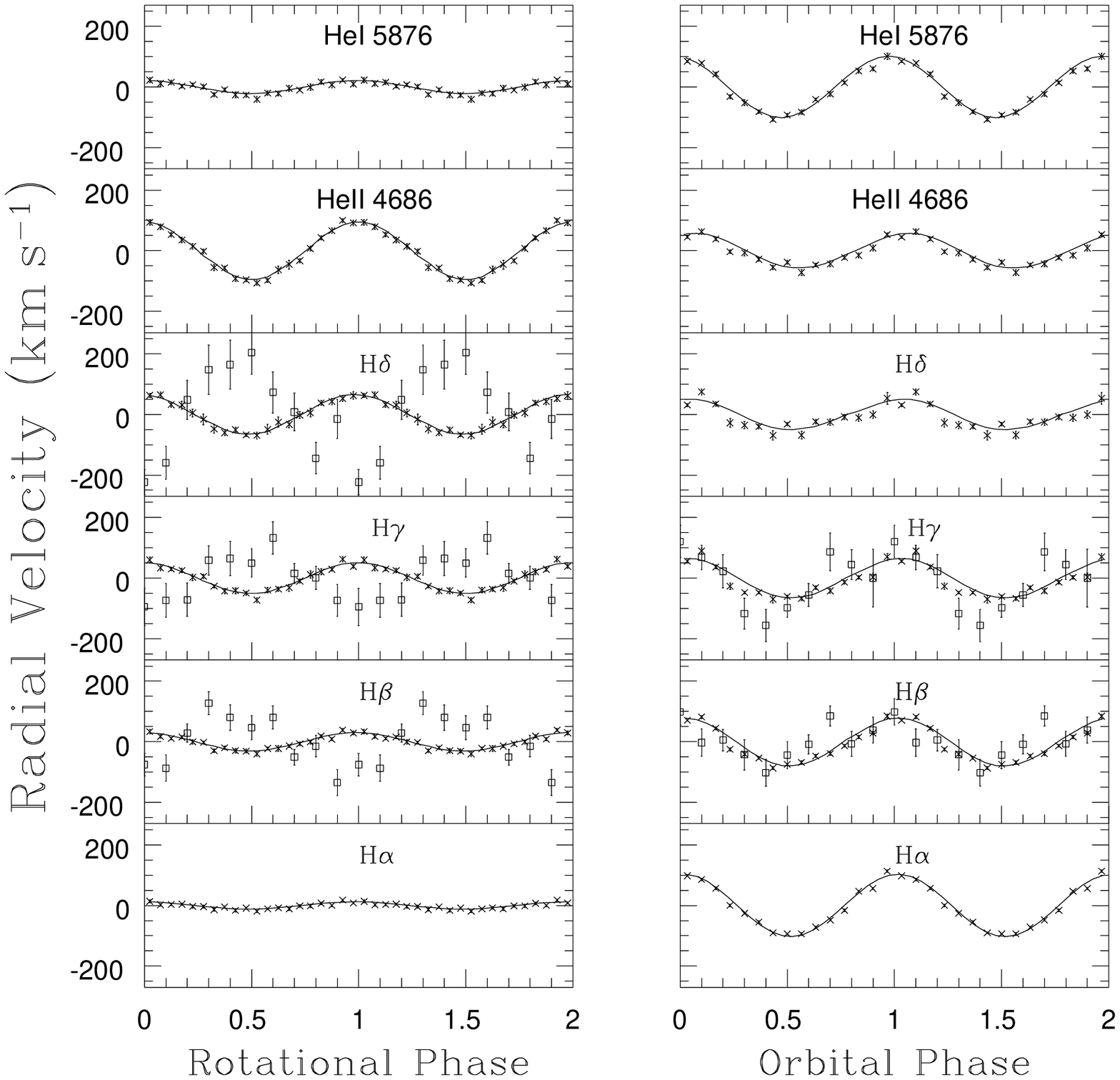}}
\caption{ {\em Left:} The radial velocity curves of Balmer 
emissions and He\,II $\lambda$ 4686 and He\,I $\lambda$ 5875  
together with their best fit two-sine  function at the spin and 
orbital periods. Note the excursion at the 
beginning of the second night observed in the blue spectra. The arrow 
marks the H$_{\alpha}$ radial velocity measurement on the red 
spectrum acquired just after the first sequence of blue spectra (see text).
{\em Center:} The folded spin radial velocity curves of the same 
emission lines in 20 phase bins detrended from the orbital  modulation 
together with their best fit sinusoidal function. {\em Right:} Folded 
orbital radial velocity curves in 20 phase bins 
detrended from the  rotational variability together with their best 
sine fit.  The radial 
velocities of the  absorption components in Balmer lines as measured on 
the 10 binned phase folded spectra at the spin and orbital periods are 
also reported  with open squares and also showing larger error bars. 
No orbital modulation could be measured in  the absorption of H$_{\delta}$ 
line (see text). $\gamma$ velocities have been subtracted.}\label{fig5}
    \end{figure*}

\subsection{Radial velocities of emission lines}

The radial velocity curves of emission lines were then fitted
with a composite two-frequency sinusoidal function plus a constant to 
account for the orbital and rotational variabilities and 
$\gamma$ velocity respectively.  Their best fit sinusoidal functions 
are shown in  Fig.\,\ref{fig5}.  We note 
that at the beginning of the second night
the radial velocities of all emission lines show a blue-to-red excursion
(not included in the timing analysis).
This is not observed on the first night where a small overlap of 
orbital phases occurs (i.e. between 0.84-0.89).
We checked on O\,I ($\lambda$ 5677) sky-line the stability 
of wavelength calibration of the blue spectra in the time interval of
the observed drift.  The changes amount to 10\,km\,s$^{-1}$, too small 
compared to the large excursion of $\sim$ 300\,km\,s$^{-1}$. 
Since the red (Grism 6) spectra overlap the H$_{\alpha}$ region, we also 
measured the  radial velocity of this line on the first red spectrum 
acquired on that 
night which shows a consistent excursion as measured on the blue spectra.
This measure is marked as an arrow in Fig.\,\ref{fig5}. 
Hence, the  observed blue-to-red excursion  of radial velocities  
appears to be real. However, given the very small overlap in orbital 
phases between the first and second night which does not cover 
the extreme red excursions in the radial velocities, it remains unclear 
whether this behaviour is due to a transient event or it is related to a 
regular occurrence at these orbital phases. Remarkably, the 
line and the continuum fluxes are not affected. Indeed,  differently from 
the radial velocities, neither the  EWs nor FWHMs of the same lines as well 
as the continuum  show any anomalous behaviour.

\begin{figure}[h!]
\centering
\includegraphics[width=7cm,angle=-90]{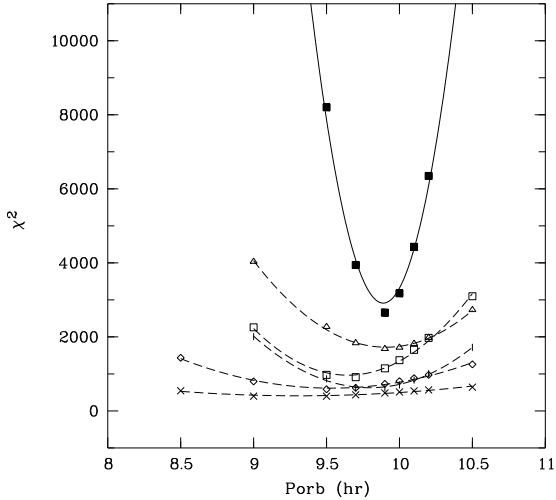}
\caption{ The $\chi^2$ periodogram of the radial velocities of emission 
lines of H$_{\alpha}$ (filled squares), H$_{\beta}$ (open squares), 
H$_{\gamma}$ (open diamonds), H$_{\delta}$ (diagonal crosses), He\,I 
(crosses) and He\,II (open triangles). The best value is found for 
H$_{\alpha}$ line.}
         \label{fig6}
\end{figure}

The sinusoidal fits to the radial velocities of emission lines give 
consistent spin periods within error bars for all lines. We then quote  
for the spin period the following ephemeris for the time of the maximum 
redshift:
$\rm HJD_{max}^{\omega}$=2452787.33751(1)+0.0080209(7)\,E, as derived from  
He\,II  line which displays the largest spin modulation and provides the 
best $\chi^2$ fit.  The derived  spin 
period $\rm P_{\omega}$= 693.01$\pm$ 0.06\,s  greatly refines  the X-ray 
and  optical values obtained  by Haberl et al. 
(\cite{Haberl02}).
In order to derive the orbital ephemeris we have detrended  the 
spin  variability from the radial 
velocities  and  performed a $\chi^2$ 
search in the same emission lines. The most accurate determination is 
obtained for the strongest H$_{\alpha}$ line, which also shows the
largest modulation, the other lines showing shallower $\chi^2$ 
minima (see Fig.\,\ref{fig6}).  The minimum $\chi^2$  corresponds to a 
period of 9.87$\pm$ 0.03\,hr which gives the following orbital ephemeris
for the time of maximum redshift:
$\rm HJD_{max}^{\Omega}$=2\,452786.7561(8)+0.411(1)\,E.

With the above ephemerides we folded the radial velocities 
at the  two periods in 20 phase bins as shown in the 
right panels of Fig.\,\ref{fig5}, previously  prewhitening them  from 
the spin and orbital variations respectively. As seen in the Fourier 
analysis, the  largest amplitude of the spin 
modulation is observed in 
He\,II  and in H$_{\delta}$, while the smallest spin 
variation is observed in H$_{\alpha}$,  and vice versa for 
the orbital variation (i.e. the amplitude increases with wavelength). 
The emission line radial velocity amplitudes,
$\gamma$ velocities and phases, as derived from the two-sine fits
are reported in Table~\ref{radvel}.

\subsection{Radial velocities of Balmer absorption lines}

To search for velocity modulations of the Balmer absorption components,
we also folded the spectra at the spin and orbital periods, again 
excluding  those affected by the above mentioned  blue-to-red 
excursion, and fitted
the Balmer lines except  H$_{\alpha}$ with two Gaussians to account for 
the absorption and emission components. The radial velocity curves  
were then  fitted with a sinusoid plus a constant accounting for the 
$\gamma$ velocity.  
 As far as the orbital radial velocity curve is concerned, only the 
absorption component of H$_{\gamma}$  gives a similar amplitude and
$\gamma$ velocity  as observed in the emission component but the 
 phase of maximum redshift occurs before that of the emission (see 
Table~\ref{radvel}).  H$_{\beta}$ gives an 
unacceptable large $\gamma$ velocity though a similar amplitude as the 
emission component. H$_{\delta}$ does not show any orbital radial 
velocity  modulation  possibly reflecting the behaviour (smaller 
amplitude with respect to the other Balmer lines) seen in the  
orbital radial velocity  of the emission component in this line.
 As for the 
spin radial velocity curve,  H$_{\gamma}$ absorption 
component  has a similar $\gamma$ velocity as the emission, but 
it is anti-phased, i.e. the maximum redshift occurs at the maximum 
blue-shift of  the emission component. Again, the $\gamma$ velocites of 
H$_{\delta}$ and H$_{\beta}$ absorption components are unacceptable, but 
their  amplitudes and  phases are similar to that of H$_{\gamma}$,
and similarly 180 degrees out of phase with the emission radial 
velocities.
The radial velocities of the absorption components are also shown in  
Fig.\,\ref{fig5}. 
A word of caution is needed for the $\gamma$ velocities, since these
result from the  combination of the systemic velocity and the 
$\gamma$ velocity  of the spin emitting region (probably 
 different for the different lines) which cannot be separated in our 
sinusoidal fits.  This might be even more crucial for the 
absorption components, which give a 
 large discrepancy in the $\gamma$ velocity values, 
if they originate in a further different region  
with respect to that of emission lines. Hence we refrain from 
interpreting 
these results.

   \begin{table*}[t!]
      \caption{Radial velocities in RX\,J1548. }
         \label{radvel}
     \centering
\begin{tabular}{l c c c c c c }
            \hline \hline
            \noalign{\smallskip}
Line &  $\gamma_{\rm em}^{\star}$ &  K$_{\rm em}^{\star}$ & 
Phase$_{\rm em}^{\star \dag}$ &
$\gamma_{\rm abs}^{\star \star}$ &  K$_{\rm abs}^{\star \star}$ & 
Phase$_{\rm abs}^{\star \star \dag}$\\    
     & km\,s$^{-1}$ & km\,s$^{-1}$ & & km\,s$^{-1}$ & km\,s$^{-1}$ &  \\
 \noalign{\smallskip}
            \hline
            \noalign{\smallskip}
Orbital Period  & & & & & & \\
            \noalign{\smallskip}
            \hline
H$_{\delta}$ &  -58(4) &  54(4) &  0.031(2) & - & -  & - \\
H$_{\gamma}$ &  -60(2) & 67(3)  & 0.025(2) & -58(13)  & 118(19) & -0.07(2)\\
H$_{\beta}$  &  -63(2) & 80(3)  & 0.014(1) & -540(14)  & 62(20) & -0.19(5)\\
H$_{\alpha}$ &  -49(4) & 103(2) &  0.000(1) &          &        & \\
He\,II $\lambda$ 4686  & -58(2) & 58(3)  & 0.046(3) &  &        & \\
He\,I $\lambda$ 5875  & -50(2) & 102(3) & -0.027(2) &  &     &    \\
     \noalign{\smallskip}
            \hline
            \noalign{\smallskip}
Spin Period & & & & & & \\
            \noalign{\smallskip}
            \hline
H$_{\delta}$ &  &  66(4)  & 0.025(2) & 200(25) & 190(33)  & -0.45(3)\\
H$_{\gamma}$ &   & 51(3)  & 0.030(2) & -64(11)  & 89(17) & -0.51(2)\\
H$_{\beta}$  &   & 30(3)  & 0.050(2) & -545(14)  & 100(18) & -0.41(2)\\
H$_{\alpha}$ &   & 12(1) & 0.002(4)  &           &         & \\
He\,II $\lambda$ 4686 &  & 95(3)  & 0.000(1) &    &     &      \\
He\,I $\lambda$ 5875  &   & 22(3) & -0.043(5) &   &     &      \\
     \noalign{\smallskip}
            \hline
\end{tabular}
~\par
\begin{flushleft}

$^{\star}$ Radial velocity curve parameters of emission lines as derived
from fitting  with  a composite two-sine 
(orbital and spin) plus a constant function the  radial velocities on 
the individual spectra.\par

$^{\star \star}$: Radial velocity curve parameters of absorption components 
as derived from fitting  with a single sine plus a constant function
the radial velocities on the phase folded spectra at the spin 
 and orbital  periods respectively.\par

$\dag$ Phase of maximum redshift using the orbital and spin 
ephemerides quoted in the text.

\end{flushleft}
\end{table*}

\subsection{Radial velocities of  the secondary star}

\begin{figure}[h!]
\centering
\includegraphics[width=7cm,angle=0]{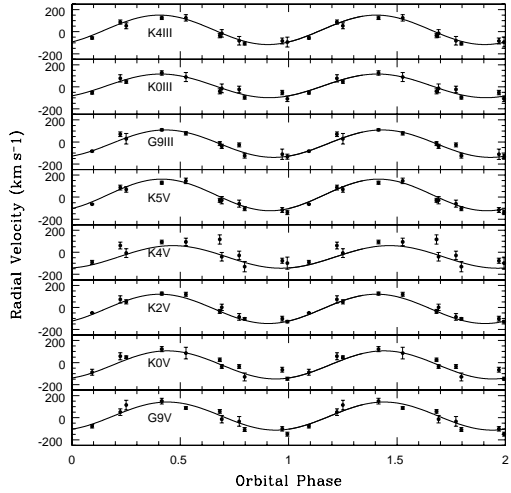}
\caption{ Radial velocity curves obtained from cross-correlation of
emission and telluric line free portion of the spectrum of RX\,J1548 with
a set of late type template star spectra.}
          \label{fig7}
\end{figure}

The red spectra containing several weak absorption features
 ascribed to the donor star were used to determine radial 
velocity variations. 
We used the range 5910-6520\,$\AA$, free from  emission lines and 
masking the telluric line at 6280\,$\AA$ and cross-correlated with a set 
of
late type star template spectra covering spectral types between G9 and 
K5 and  luminosity class V and III. Since the few template stars observed 
with Grism 6 do not cover 
the whole range of spectral types and luminosity class, we  
used further spectra taken from the Jacoby library  (Jacoby et al. 
\cite{jacobyetal84}).
The cross-correlation functions were then used to obtain the radial 
velocity curves shown in Fig.\,\ref{fig7}. These curves were fitted 
with a sinusoid and the results are reported in Table\,\ref{secpar}. 
Similar  quality fits are  found for  a K4III  
($\chi^2_{red}$=1.01),  for a K2V 
($\rm \chi^2_{red}$=1.05)  and for a K5V star ($\rm 
\chi^2_{red}$= 1.07). 
Amplitudes in 
these fits are consistent within  their 1-$\sigma$ errors. The $\gamma$ 
values are not  reported since 
Jacoby library spectra have unknown epoch of observations. Worth 
noticing is the phasing of maximum redshift, which  precedes the 
emission line maximum blue shift by $\sim$0.1, but it is in phase with the 
maximum blueshift of H$_{\gamma}$ absorption.  We 
also performed similar cross-correlation 
on the same range in the blue spectra binned in 10 orbital phases and
found similar results. Hence, the radial velocities derived from 
cross-correlation unambiguously map the donor star orbital motion with 
amplitude in the range 100-143\,km\,s$^{-1}$, though these do not provide 
strong constraints on its spectral type.
 Roche geometry and Kepler's law for a binary orbital period of 
9.87\,hr gives a secondary mean density 
of 0.78$\rho_ {\odot}$, much larger than that of giants (Schmidt-Kaler 
\cite{schmidtkaler}) and  subgiants (Dworak \cite{dworak}) and hence
a main sequence star is adopted (but see Sect.\,5).

   \begin{table}[h]
      \caption{Orbital radial velocity solutions for the secondary star in 
RX\,J1548.}
         \label{secpar}
     \centering
\begin{tabular}{l c c c}
            \hline \hline
            \noalign{\smallskip}
Sp. Type  &  K$_{sec}$ &  Phase$^1$ & $\chi^2_{red}$  \\
          & km\,s$^{-1}$ &        & \\ 
\noalign{\smallskip}
            \hline
            \noalign{\smallskip}
G9\,V &  127$\pm$15 & 0.44(3) & 3.10\\
K0\,V &  127$\pm$16 & 0.44(2) & 2.85\\
K2\,V &  133$\pm$7  & 0.41(1) & 1.05\\
K4\,V & 102$\pm$16 & 0.46(5) & 2.07\\
K5\,V & 143$\pm$9 &  0.41(1) & 1.07\\
G9\,III & 124$\pm$10 & 0.43(2) & 2.34\\
KO\,III & 107$\pm$13 & 0.41(3) & 1.70\\
K4\,III & 133$\pm$9 & 0.41(1) & 1.01\\
     \noalign{\smallskip}
            \hline
\end{tabular}

~\par
\begin{flushleft}
$^1$: Phase of maximum redshift using the orbital ephemeris quoted in the 
text.
\end{flushleft}

\end{table}


\section{Discussion}

Our new spectroscopy on RX\,J1548 allows us to precisely determine the 
orbital period of this binary, $\rm P_{\Omega}$=9.87$\pm$0.03\,hr, which 
locates it at the far end of the observed orbital periods of IPs. 
RX\,J1548 thus joins 
V\,1062 Tau (9.95\,hr),  AE\,Aqr (9.88\,hr) as a third system just
outside the  claimed IP period gap, with only RX\,J2133.7+5107 
(Bonnet-Bidaud et 
al. \cite{bonnet-bidaud05}) inside it,  and as a fifth member of 
the long  period IP 
group when including the old nova GK Per (47.9\,hr) and the  recently 
discovered IP RX\,J173021.5-0559 (15.4\,hr) (G\"ansicke et al. 
\cite{gaensicke05}).  While the previously known long
period IPs show strong variability (low states, outbursts or flares),   
RX\,J1548 and RX\,J173021.5-0559  are not known yet to show  peculiar 
behaviour. For RX\,J1548 we can only quote 
a  possibly transient variability in 
the emission line radial velocities and a long term  change in the flux 
level of  0.4\,mag on a five years timescale. Clearly long term monitoring 
of this southern system is needed to characterize its 
 long-term behaviour.

The 693\,s modulation is detected in the 
radial velocities of emission lines thus  greatly refining 
($\rm P_{\omega}=693.01 \pm$0.06\,s) the 
previously detected X-ray and optical photometric periods and confirming 
it as the spin period of the accreting WD. Though the data might be 
partially affected by slit losses a strong spin variability does not seem 
to be present in the  continuum during  our observing run in 2003, 
contrary to the photometric observations of Haberl et 
al.(\cite{Haberl02}).   The  presence of the spin period in the 
radial velocities and the absence of any beat modulation confirms that 
the WD accretes  via a disc. 

EWs of Balmer and He\,I emission lines show weak indication of spin 
modulation 
which is absent in  the EWs of He\,II line. On the other hand 
the amplitudes of the spin radial velocity curves are largest
for He\,II and smallest for H$_{\alpha}$. All this might indicate that the
emission lines  originate in an optically thin region of the accretion 
flow onto the WD, with  He\,II forming in a more localized region.
The lack of spectral coverage of the Balmer jump 
hampers  an estimate of the optical thickness of the spin modulated line 
forming region.  
 
\subsection{The stellar components and system parameters}

The donor star in RX\,J1548 is detected from red 
absorption features in  its spectrum and their identification suggests a  
spectral type between K0-K3\,V. 
Using the empirical CV period-spectral type relation of 
Smith \& Dhillon (\cite{smithdhillon}), for an orbital period of 9.87\,hr a 
G9$\pm3$ is expected, consistent with our findings.
We then adopt for secondary star in RX\,J1548 a K2$\pm$2\,V 
type. From the ratio of EWs  
of absorption lines at 6162\,$\AA$ and at 6490\,$\AA$ and those of 
K0-K4 type standards, the contribution of the stellar flux   
to the total observed flux at 6100\,$\AA$ continuum is estimated to be
20$\pm$3$\%$. For a K2$\pm$2\,V star the  absolute magnitude 
is in the range  
$\rm M_V$=5.9-7.0\,mag.  This  yields a distance in the range 
770-1200\,pc. X-ray data 
give an hydrogen column density $\rm N_H=1.47\pm 0.08\times 
10^{21}\,cm^{-2}$, 
which  provides an upper limit to the interstellar extinction in the 
direction of the source. Haberl et al. (\cite{Haberl02}) estimate then 
$\rm E_{B-V}\sim 0.25$ thus  lowering the distance to 540-840\,pc. 
 This estimate of the distance can be used to provide a constraint on 
the WD flux contribution. For 
distances in the range 770-1200\,pc, the WD flux is only 
1-4$\%$ of the observed flux at H$_{\gamma}$ continuum. 
This amounts to 3-4$\%$ when distance is 540-840\,pc,  taking into
account the extinction. This implies that the Balmer absorption 
features cannot be safely ascribed to the WD atmosphere. 
The orbital radial velocity curve of H$_{\gamma}$ absorption
is out of phase by most exactly 180$^o$ with that of the secondary star, 
thus indicating that the absorption line forming region is located very 
close to the WD surface. This conclusion is also consistent with the fact 
that the H$_{\gamma}$ absorption  curve is also closely in phase with the 
emission line orbital radial velocities, but  preceding them by 0.1 
in phase. Moreover, its spin radial velocity curve   is 
anti-phased with respect to that of  emission lines, suggesting  
 an optically thick region  
offset by 180$^o$ from the main accreting pole where the  emission lines 
are formed.  This region could be  the accretion 
flow onto the opposite pole. However, this possibility does 
not explain the  larger widths of the Balmer absorption components with 
respect to the emission components. An origin in the WD atmosphere is 
difficult not only because the  estimated WD  contribution to the total 
observed flux is negligible, but  also   because no  radial velocity 
motion at the WD rotational period is  expected. 
Though an origin in the disc cannot  be excluded, as this would produce 
an orbital radial velocity curve with similar amplitudes and phases as 
observed in the emission lines (especially 
taking into  account that the  system  has a very long period and probably 
it is a low  inclination binary (see below)), a spin radial  velocity 
curve  cannot be  accounted by a disc and the system is not accreting at a 
high mass rate (see sect.\,5.2), needed to produce optically thick 
absorption lines. A further alternative possibility is 
that the Balmer absorption components originate in an accretion halo 
similarly to what observed in some Polar systems like V834\,Cen, MR\,Ser 
and EF\,Eri (Achilleos et al. 
\cite{Achilleos92}, Schwope et al. \cite{Schwope93}, Wickramasinghe \& 
Ferrario \cite{Wickra00}). In these Polars broad Zeeman-shifted 
absorption  components of  Balmer lines are seen during high 
mass accretion states and during the bright cyclotron-dominated phases and 
arise from a cool halo of unshocked gas surrounding the accretion shock 
close to the WD surface. The magnetic accretion halo scenario could 
account 
for the large widths of absorption components in RX\,J1548 since 
the FWHMs of H$_{\beta}$  and H$_{\gamma}$ absorptions, if interpreted as 
unresolved $\sigma$ and $\pi$ Zeeman components,  
would  imply a magnetic  field strength $\sim$2-4\,MG. However,
at these magnetic field strengths these components should be  
clearly resolved in  H$_{\alpha}$ line. If the 
accretion halo is optically thick, as seen in higher Balmer lines, the 
absorption in H$_{\alpha}$ should be stronger because the ratio of the
oscillator strengths of the $\pi$ components of H$_{\alpha}$ and
H$_{\beta}$ is $\sim$8 (Wickramasinghe et al. \cite{Wickram87}). But no 
absorption features are observed close to H$_{\alpha}$ in either
the average spectrum or the spin phase-resolved spectra. 
The FWZI of H$_{\alpha}$ emission line instead would imply a magnetic field 
lower than 1\,MG, to mask the Zeeman absorption components. Furthermore,   
phase-dependent redshifts of these components were observed in some of 
these Polars (maximum  redshift when viewing along the magnetic field lines)
due to Doppler effect in the free-falling accretion flow and/or by 
variations in the effective field strength. An anti-phased spin 
behaviour of the absorption features with respect to 
the emission lines as detected in RX\,J1548 would then imply  
a complex accretion halo structure at the secondary pole. Finally, if  
an accretion  halo is present in 
RX\,J1548, this system should reveal a relatively strong polarized emission.
Hence, while  different scenarios could be envisaged, they do not 
simultaneously explain the observed Balmer absorption characteristics and 
a satisfactorly explanation should await future spectropolarimetric 
observations.

\begin{figure}[h]
   \centering
\includegraphics[width=8.cm,angle=0]{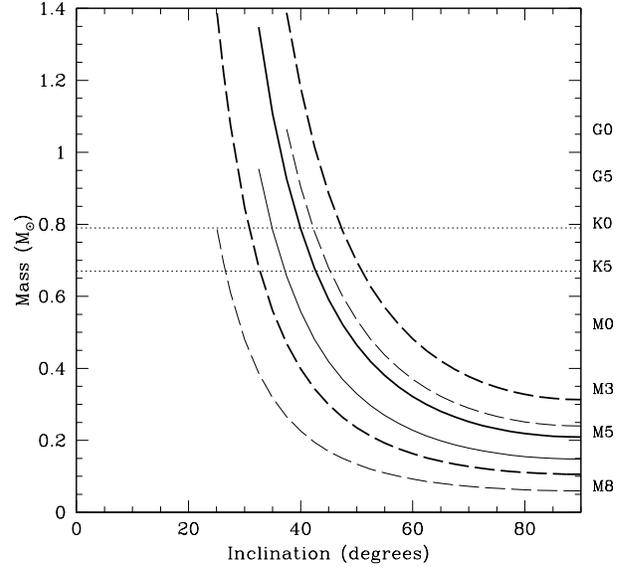}
\caption{The  WD (thick curves) and secondary star 
(thin curves) masses versus inclination 
adopting a mass ratio $q=0.65\pm$ 0.12. The dashed upper and lower curves 
correspond to the uncertainties in the mass ratio.  The lower limit
on inclination is set by the Chandrasekhar limit for the WD mass.  The 
main sequence mass-spectral type relation is also reported on the right
side. Horizontal dotted lines mark the adopted range of 
spectral types for the secondary star. }
\label{fig8}
    \end{figure}

Though caution has to be taken to ascribe emission line 
radial velocities as tracers of the WD orbital motion, as they 
can originate in different regions within the binary system, the fact that 
the Balmer absorption amplitudes are,  to within error, consistent 
with  the emission lines gives support to their use in the radial 
velocity solutions for mass determination. We then adopt conservatively
the wide range  $\rm K_{WD}$ = 54-102\,km\,s$^{-1}$.  
For the K2$\pm$2\,V secondary we adopt: $\rm 
K_{sec}$=102-133\,km\,s$^{-1}$  and 
then the mass ratio is $q=\rm M_{sec}/M_{WD}$=0.53-0.77.
This ratio limits the WD and secondary star 
masses  in the mass-inclination plane, giving $\rm M_{WD}$\,$sin\,i^{3}$ = 
0.21$\pm$0.10\,M$_{\odot}$ and  $\rm M_{sec}$\,$sin\,i^{3}$= 
0.14$\pm$0.09\,M$_{\odot}$,  as depicted in Fig.\,\ref{fig8}. 
The condition 
that the WD cannot be larger than the Chandrasekhar value implies that
$i>25^o$ and since no eclipses are observed, $i<75^o$. 
Although the dynamical solution gives a wide secondary mass range, 
the secondary spectral type (K0 to K4), adopting the main sequence 
mass-spectral type relation (Schmidt-Kaler \cite{schmidtkaler}),  is used 
to define the two horizontal lines in Fig.\,\ref{fig8}. These limit the 
component masses
to $\rm 0.89\,M_{\odot} \leq M_{WD} < M_{Chandr.}$ and $\rm 
M_{sec} \sim 0.68-0.79\,M_{\odot}$ and restricts the inclination 
in the range $25^o < i \leq 45^o$.  From the mass ratio,  the 
secondary star Roche lobe dimension can be determined ($\rm 
R_{lobe}$/$a = 0.47\,[q/(1+q)]^{1/3}$), giving  $\rm 
R_{lobe}$=$0.33-0.36\,a$, where $a$ is the binary separation.
For the inclination under consideration and using the 
secondary star radial velocity amplitude, this gives $\rm 
R_{lobe} \sim 0.7-1.3\,R_{\odot}$.  A comparison with the mass-radius 
relation for low mass
main sequence stars (Baraffe et al. \cite{baraffe}), indicates that 
the secondary star on average has a radius  larger 
than that of a main sequence star of comparable mass. 
As discussed by Beuermann et al. (\cite{Beuermannetal}), 
donor stars in the long period CVs are 
typically cooler than ZAMS stars with the same  mass
 and their position  
in the spectral type-orbital  period plane (cf. their Fig.\,5) 
indicates evolved secondaries. 
The derived orbital period  and the secondary spectral type locate  
RX\,J1548 in that diagram close to evolved models of stars at the 
end of core hydrogen burning, thus  
suggesting that the donor has undergone substantial nuclear 
evolution. Furthermore,
Kolb et al. (\cite{kolbetal01}) demonstrated that  the mass of a 
ZAMS star with the same spectral 
type as the  donor  should be considered as an upper limit, while
the lower limit depends on the still uncertain mixing length parameter.  
For spectral types earlier than K6, however, a lower limit can be 
set, which decreases the ZAMS mass by $\sim$ 0.3\,$M_{\odot}$ for a  
spectral type between K0 and K4. 
Hence, because RX\,J1548 appears to  host an 
evolved secondary, we conservatively adopt  $\rm 0.4 \leq M_{sec} 
\leq 
0.79\,M_{\odot}$ and then $\rm M_{WD} \geq 0.5\,M_{\odot}$
and $25^o < i \leq 58^o$.

\subsection{Magnetic accretion}

With our period determination,  the spin-orbit ratio $\rm 
P_{\omega}/P_{\Omega}$ of 0.02 indicates a high 
degree of asynchronism. For comparison, the other IPs with similar 
orbital 
periods have very different  degrees of asynchronism:  
AE\,Aqr with the extreme $\rm P_{\omega}/P_{\Omega}$=0.00093 and  
V1062\,Tau with the typical value found in most IPs $\rm 
(P_{\omega}/P_{\Omega}$=0.10) (see also King \& Lasota \cite{kinglasota}).
RX\,J1548 is thus an intermediate system among the two, whose  
 rotational equilibirum needs to be investigated. 
We thus evalute the corotation radius  at which the magnetic
field rotates at the same rate as the Keplerian frequency $\rm 
R_{co} = (G\,M_{WD}\,P_{\omega}^2/4\,\pi^2)^{1/3}$. For a WD with
$\rm 0.5\,M_{\odot} \leq M_{WD} < 1.4\,M_{\odot}$ and $\rm 
P_{\omega}$=693\,s,  $\rm 
R_{co}=0.91-1.29\times 10^{10}$\,cm. The condition for accretion will 
require $\rm R_{mag} \leq R_{co}$ where 
$\rm R_{mag}= 5.5\times 
10^{8}\,(M_{WD}/M_{\odot})^{1/7}\,R_{9}^{-2/7}\,L_{33}^{-2/7}\,\mu_{30}^{4/7}$\,cm, 
where $\rm R_{9}$ is the WD radius (in units of $10^{9}$\,cm), $\rm 
L_{33}$ 
is the luminosity (in units of $\rm 10^{33}\,erg\,s^{-1}$) and $\mu_{30}$ 
is 
the magnetic moment (in units of $\rm 10^{30}$\,G\,cm$^3$). 
Haberl et al. (\cite{Haberl02}) derive an unabsorbed flux of 3.3$\times 
10^{-11}\rm \,erg\,cm^{-2}\,s^{-1}$ for the optically thin emission 
component, and a bolometric blackbody flux of $\rm 1.5\times 
10^{-11} \,erg\,cm^{-2}\,s^{-1}$ which allow to estimate the  
accretion luminosity  $\rm L_{acc}=5.74\times 10^{31}\,d^{2}_{100}$erg\,s$^{-1}$, where $\rm 
d_{100}$ is the distance in units of 100\,pc. 
The condition for accretion at spin equilibrium, using the above WD mass 
range,  then requires a magnetic moment $\mu < 4.9\times 
10^{31}\,d_{100}$\,G\,cm$^3$. 
For a spin-to-orbit period ratio of 0.02 and an orbital period of 9.87\,hr 
the magnetic model of spin equilibria by Norton et al. 
(\cite{norton04}) predicts a magnetic moment $\rm \mu \sim 1.5\times 
10^{33}\,G\,cm^3$, for $q=0.5$, which,  is larger than what estimated 
above even allowing a 40$\%$ variation of magnetic moment for 
$0.1<q<0.9$. For such high 
magnetic moment the condition of spin equilibrium would imply an accretion 
rate $\rm >7\times 10^{17}\,g\,s^{-1}$.  However, from the  luminosity 
obtained  from X-rays, we estimate a  mass  accretion rate
$\rm \dot M \leq 7\times 10^{14}\,d^2_{100}\, g\,s^{-1}$, which is also 
much lower than the secular
value for a CV with a 9.87\,hr orbital period (Norton et al. 
\cite{norton04}). Hence RX\,J1548 is probably far from equilibrium
and the WD appears to possess a weak magnetic 
field  strength. This was also noticed independently by Haberl et al. 
(\cite{Haberl02}) as this IP possess a soft X-ray component with a 
moderate ratio (46$\%$) of blackbody-to-hard X-ray luminosity  when 
compared to other soft X-ray/high magnetic field IPs (de 
Martino et al. \cite{demartino04}).

\section{Conclusions}

Our optical spectroscopy of RX\,J1548 has allowed to refine 
the orbital  and WD spin periods detected by Haberl 
et al. (\cite{Haberl02})  and to derive information on this 
binary  as follows:

\begin{itemize}

\item  The orbital period of 9.87\,hr locates RX\,J1548 just outside the
IP period gap, thus joining other four long period systems (AE\,Aqr, 
V1062 Tau, RX\,J173021.5-0559 and GK\,Per). Because of lack of long term 
monitoring,  we are unable to assess whether RX\,J1548 also shows a 
peculiar optical behaviour as observed in the well monitored AE\,Aqr, 
V1062 Tau and GK\,Per. We only 
detect a  strong  variability in the radial 
velocities of emission 
lines and a long term luminosity variation on a five years timescale.\\

\item The 693\,s spin period is detected in the radial velocities of 
emission lines and of the absorption of Balmer lines.  Though 
not truly spectrophotometric, our data do not reveal a strong 
spin or  beat periodicity in the continuum contrary
to  previous photometric observations. \\

\item From weak absorption features in the red portion of the spectrum we 
identify the spectral type of 
the secondary star as a K2$\pm$2\,V. 
The orbital period and the spectral type suggest that the donor star 
in RX\,J1548 has undergone nuclear evolution as also found in other 
long period CVs.\\

\item The amplitude of the orbital radial velocites of emission and weak
absorption lines allows us to determine the mass ratio $q=0.65\pm 0.12$.
From the secondary star spectral type, allowing for uncertainties in the
mass-spectral type relation for CV donors, we find  $\rm 
M_{WD} \geq 0.5\,M_{\odot}$ and $\rm M_{sec}=0.4-0.79\,M_{\odot}$ and
a binary inclination  $25^o < i \leq 58^o$.\\

\item We estimate the distance in the range 540-840\,pc which 
limits to only 3-4\,$\%$ the contribution of the WD to the observed 
flux
at H$_{\gamma}$ continuum and to 20-25\,$\%$ the contribution of the 
secondary star at 6100\,$\AA$. This implies that the observed absorption 
features underneath Balmer emissions are not due to the WD atmosphere.
RX\,J1548 is not accreting at a high rate which does  not favour  
an optically  thick disc as instead observed in high transfer
rate systems such as  dwarf novae during outbursts or in SW Sex stars. 
The observed orbital and spin radial velocities of H$_{\gamma}$ 
absorption instead suggest an origin  in the accretion flow close to 
the WD surface which needs further investigation via spectropolarimetric 
observations.\\ 

\item The  spin-to-orbit period ratio $\rm P_{\omega}/P_{\Omega}$=0.02 
suggests  that this system is probably far from equilibrium. Also, the WD  
appears to be weakly magnetized with $\rm B < $ 2\,MG,  
much lower than typical values found in the other known soft X-ray/highly 
polarized IPs. Polarimetric observations are then needed to clarify whether 
this  system really contains a low field WD.

\end{itemize}

 \begin{acknowledgements}
DdM acknowledges financial  support by the Italian Ministry of 
University and Research (MIUR). BTG was supported by a PPARC Advanced 
Fellowship. 
\end{acknowledgements}

\end{document}